\documentclass[twocolumn,preprintnumbers,amsmath,floatfix,
  aps,pra,superscriptaddress]{revtex4} 

\usepackage{amsmath}
\usepackage{graphicx} 
\usepackage{amssymb}

\makeatletter

\begin{document}

\title{Bound on quantum computation time: Quantum error correction in
  a critical environment}

\author{E. Novais}
\affiliation{Centro de Ci\^encias Naturais e Humanas, Universidade
  Federal do ABC, Santo Andr\'e, S~ao Paulo, Brazil}

\author{Eduardo R. Mucciolo}
\affiliation{Department of Physics, University of Central Florida, Box 162385,
  Orlando, Florida 32816, USA}

\author{Harold U. Baranger}
\affiliation{Department of Physics, Duke University, Box 90305,
  Durham, North Carolina 27708-0305, USA}

\date{April 20, 2010}
\begin{abstract}
We obtain an upper bound on the time available for quantum computation
for a given quantum computer and decohering environment with quantum
error correction implemented. First, we derive an explicit quantum
evolution operator for the logical qubits and show that it has the
same form as that for the physical qubits but with a reduced coupling
strength to the environment. Using this evolution operator, we find
the trace distance between the real and ideal states of the logical
qubits in two cases. For a super-Ohmic bath, the trace distance
saturates, while for Ohmic or sub-Ohmic baths, there is a finite time
before the trace distance exceeds a value set by the user.
\end{abstract}
\maketitle

\noindent
\emph{Introduction}-- All physical implementations of quantum
information processing face the inexorable reality of being embedded
in an environment that causes decoherence \cite{stamp}. There are many
strategies for dealing with this fact \cite{NielsenChuang}, quantum
error correction (QEC) being one of the most versatile \cite{QEC}. As
QEC will likely be used in any large-scale implementation of quantum
information processing, it is important to define and to quantify its
limits.

There have been several recent advances in understanding the limits of
fault tolerant quantum computing \cite{Lidarbook}. Part of this effort
has been on proving ``resilience'', the notion that any desired
accuracy of quantum computation may be attained by concatenating
levels of the QEC code \cite{ref4}. In particular, arguments for
resilience in correlated environments have been constructed either by
using techniques based on sums over faulty paths
\cite{B-T,latestPreskill} or by reducing the problem to an almost
stochastic one through scaling \cite{NB06}. In this Rapid
Communication, we focus on a related question: Given a certain quantum
computer and an environment, what is an upper bound on the time
available for computation? We provide an answer for a broad class of
environments using a Hamiltonian formulation, including those where
correlation effects are induced by gapless modes (i.e.\ critical
environments).

The main results of our argument are as follows. First, the
coarse-grained quantum evolution of logical qubits is essentially the
same as that of physical qubits, up to a renormalized coupling to the
environment. Thus QEC maps our generic environment-computer
interaction model onto itself, a property which has proven useful for
other ways of suppressing decoherence such as dynamical decoupling
\cite{DD+FT}. Second, we use this coarse-grained quantum evolution to
find the maximum time available for computation, as assessed by the
trace distance between the real and ideal states of the computer.
There is a regime where computation can continue indefinitely and so
is resilient, while in other regimes the maximum time depends strongly
on the QEC code, number of logical qubits, and environment-computer
interaction.

\noindent
\emph{Environment-computer interaction and hypotheses}-- Many physical
noise cases can be directly modeled by the ``spin-boson model''
\cite{ref3}. However, as originally proposed \cite{caldeiraleggett},
this model has a much more fundamental appeal.

Let us assume that the qubits are already under the protection of an
initial strategy, such as a decoherence free subspace or dynamical
decoupling \cite{DD+DFS}. Though it deals with the dominant effect, it
is unlikely to solve the decoherence problem completely. The
inevitable residual interaction between the computer and the
environment imposes a pointer basis for the qubits, which we call the
$z$ direction of each individual qubit. Another reasonable assumption
is that the environment consists of a very large set of quantum
degrees of freedom with some intrinsic dissipative mechanism. Hence,
in the absence of the qubits, the environment will be in a local
minimum of its energy landscape. Our next (crucial) assumption is that
linear response describes the influence of the computer on the
environment. In that case, the computer-environment interaction cannot
take the environment out of its local minimum, and so we may use the
harmonic approximation to describe the quantum fluctuations of the
environmental degrees of freedom. By construction, then, the
environment-computer interaction term is linear. Thus, we use the
well-known and experimentally relevant spin-boson model to discuss the
limits on protection that QEC and fault-tolerant methods can yield.

Having thus settled the model of the strongest channel of residual
decoherence, we consider the presence of an additional weaker
transverse channel, denoted by $x$. If the first channel were absent,
the preceding argument applied to the transverse channel leads to a
model of the same form but with a much weaker coupling:
$\lambda_{x}\ll\lambda_{z}$. Hence, we describe the residual
decoherence of the quantum computer by two bosonic baths ($\hbar=1$),
\begin{equation}
H_{0}  =  \sum_{\alpha=\{ x,z\}} \sum_{\left|{\bf k}\right| \neq 0}
\omega_{\alpha,k}\, a_{\alpha,{\bf k}}^{\dagger}a_{\alpha,{\bf k}}.
\label{eq:freeHamil}
\end{equation}
The $a_{\alpha,{\bf k}}$ obey standard commutation relations, and
$\omega_{\alpha,k} = \omega_0\, (|{\bf k}|/k_0)^{z_\alpha}$, where
$k_0$ and $\omega_0$ are constants with dimensions of momentum and
frequency, respectively, and $z_\alpha$ is a dynamical exponent. The
environment-computer interaction has the form
\begin{equation} 
H_I  =  \sum_{\alpha=\{x,z\}} \sum_{\bf x} \lambda_\alpha :\!
f^\alpha ({\bf x})\!:\, \sigma_{\bf x}^\alpha,
\end{equation}
where $\sigma_{\bf x}^{x,z}$ are the Pauli matrices for the qubit at
position ${\bf x}$, $::$ stands for normal ordering, and
$
:\! f^\alpha ({\bf x})\!:  \,= \, 
(2\pi/L)^{D/2} 
\sum_{{\bf k}\neq 0} \left( u_{\alpha,k}
  e^{i{\bf k} \cdot {\bf x}} a_{\alpha,{\bf k}}^{\dagger} +
  \mbox{H.c.} \right),
\label{eq:f}
$
with $\left|u_{\alpha,k}\right|^{2} = \kappa_{0}^{-D} (\left|{\bf
  k}\right|/k_0)^{2s_\alpha}$ which defines the exponent
$s_\alpha$. Here, the environment has spatial dimension $D$ and
smallest momentum $2\pi/L$, and $\kappa_0$ is a constant with
dimensions of momentum. There is no {\sl a priori} restriction on
including a third bath; however, it would be a redundant description
of the possible errors. All the bosonic averages performed are done
with respect to the bosonic vacuum with no initial entanglement
between the computer and the bath. If there were some initial
entanglement, it could be modeled using a finite temperature in the
bosonic correlators, thus introducing an exponential scale. Since our
goal is to calculate an upper bound for the computational time, we do
not consider this case.

In conjunction with this model, we make some assumptions about the
computer and the QEC method. (i) Gates are flawless and are done much
faster than the QEC period $\Delta$. (ii) State preparation and
measurements are done perfectly. (iii) Lowest order perturbation
theory in $H_I$ is enough to describe the evolution during a QEC
step. (iv) \emph{All the syndromes indicate a non-error result}; that
is, we consider the most favorable quantum computer evolution, as any
other evolution will involve a larger leak of information to the
environment \cite{NB06}.

\noindent
\emph{Uncorrectable errors and the quantum evolution}-- The first step
of the quantum calculation is to consider the evolution operator in
the interaction picture $U_I (\Delta,0) = T_t \exp \left[ -i
  \int_0^\Delta dt\, H_I (t) \right]$ up to a time $\Delta^-$, just
before error correction is applied. QEC divides errors into classes
that can be distinguished from each other; however, within each class,
different errors are not differentiated by the syndrome. For each
logical qubit the syndrome breaks the evolution into $u(\Delta^+,0) =
\sum_{i=0}^{N-1} v_i$, where $i$ indexes the $N$ possible syndromes of
that qubit \cite{NB06}. Each one of these evolution operators has a
``good'' and a ``bad'' part: $v_i = \alpha_i + \sum_{j=1}^{3}
\beta_i^j \bar{\sigma}^j$, where $\bar{\sigma}^j$ represents a logical
error. These logical errors are uncorrectable (or ``bad evolutions''
\cite{PreskillNotes}).

Following our hypotheses, within the QEC period $\Delta$, we
approximate the evolution operator by expanding to lowest order in the
couplings $\lambda_\alpha$. (Technically, the expansion parameters are
$\lambda_{\alpha}\Delta$ and not simply $\lambda_\alpha$.) Thus, for a
single qubit,
\begin{equation}
U_I (\Delta,0) \approx 1 -i \sum_{\alpha=\{ x,z\}} \sum_{\bf x}
\lambda_\alpha \Delta :\! f^\alpha ({\bf x},0) \!: 
\sigma_{\bf x}^\alpha.
\label{eq:evol-expanded}
\end{equation}
For a code of distance $d_c$, one finds that the lowest order term
that must be kept in each logical qubit is of order $d_c$ in the
coupling to the environment. For illustration, consider the smallest
distance-3 code, namely, the 5-qubit code \cite{NielsenChuang}. At the
end of a QEC period, there are $4^5$ possible configurations for the
five qubits. They are divided into $4^2$ groups with distinct
syndromes; however, each group has $4^3$ elements that cannot be
distinguished by the code. We choose to analyze the evolution for
which all the syndromes are the ``no error'' type. This yields the
quantum evolution operator
\begin{eqnarray}
v_0 (\Delta,0) & \approx & \bar{1} + i \Delta^3 \sum_{{\bf
    x},\alpha,\beta,i,j,k} \eta_{ijk}^{\alpha\beta} \lambda_\alpha
\lambda_\beta^{2}\! \label{eq:evol-1} \\ & & \times :\!  f^\alpha
({\bf x}_i,0)\!: :\! f^\beta ({\bf x}_j,0)\!::\! f^\beta({\bf
  x}_k,0)\!: \!\bar{\sigma}_{\bf x}^\alpha, \nonumber
\end{eqnarray}
with ${\bf x}$ labeling the logical qubits, $\alpha,\beta = \left\{
x,z\right\}$, and $i,j,k = \left\{ 1,...,5 \right\}$ labeling the
physical qubits inside the logical qubit ${\bf x}$. Each coefficient
$\eta_{ijk}^{\alpha\beta}$ has two possible values, $\eta_{324}^{xz} =
\eta_{435}^{xz} = \eta_{514}^{xz} = \eta_{125}^{xz} = \eta_{213}^{xz}
= \eta_{134}^{zx} = \eta_{412}^{zx} = \eta_{245}^{zx} =
\eta_{523}^{zx} = \eta_{315}^{zx} = 1$ and zero for all other indices.

Now, we use the commutation relations of the free bosons to normal
order the evolution operator in Eq.\,(\ref{eq:evol-1}),
\begin{eqnarray}
\lefteqn{ v_0 (\Delta,0) \approx \bar{1} + i \sum_{{\bf
      x},\alpha,\beta,i,j,k} \eta_{ijk}^{\alpha\beta} \lambda_\alpha
  \Delta :\! f^\alpha ({\bf x}_i,0)\!: } & & \nonumber \\ & & \quad
\times \left[a_{\beta jk} +\ (\lambda_\beta \Delta)^{2}
  :\!f^\beta({\bf x}_{j},0) f^\beta ({\bf x}_{k},0)\!: \right]
\bar{\sigma}_{\bf x}^\alpha,
\label{eq:evol-2}
\end{eqnarray}
where $a_{\alpha ij} = (\lambda_\alpha \Delta)^{2} \sum_{{\bf k} \neq
  0} \left| u_{\alpha,k} \right|^{2} \exp \left[ -i{\bf k} \cdot ({\bf
    x}_{i} - {\bf x}_{j}) \right]$. Equation (\ref{eq:evol-2}) is
written for the 5-qubit code with no concatenation; it is
straightforward to generalize it to a larger distance or concatenated
code. For instance, the level-1 concatenated code requires 25 physical
qubits with the coefficients $\eta$ changing accordingly. In this
case, $\Delta$ includes the time needed to extract all syndromes
(including level-1 syndromes), and uncorrectable errors appear at
higher order in $\lambda_{\alpha}$.

The evolution operator for a logical qubit can be abridged by
rewriting Eq.\,(\ref{eq:evol-2}) as
\begin{equation}
v_0 (\Delta,0) \approx \bar{1} + i \Delta \sum_{{\bf
    x},\alpha=\{x,z\}} \!\!\! (\lambda_\alpha^\ast + \Gamma_\alpha)
:\!f^\alpha ({\bf x},0)\!: \, \bar{\sigma}_{\bf x}^\alpha,
\end{equation}
where ${\bf x}$ is the average position of the physical qubits
belonging to the logical qubit, $\lambda_\alpha^\ast \equiv
\lambda_\alpha \sum_{\beta,i,j,k} \eta_{ijk}^{\alpha\beta} a_{\beta
  jk}$, is the effective coupling constant, and $\Gamma_\alpha ({\bf
  x},0) \equiv \lambda_\alpha \sum_{\beta,i,j,k}
\eta_{ijk}^{\alpha\beta} (\lambda_\beta \Delta)^2 :\!f^\beta ({\bf
  x}_{j},0) f^\beta ({\bf x}_{k},0)\!:$ accounts for higher-\-order
corrections. The latter dresses the single logical qubit amplitude
$a_{\beta jk}$ by the interactions with other logical qubits. If the
spatial separation of logical qubits is at least $\Xi$ while that of
the physical qubits within a logical qubit is $\xi$, then
$\Gamma_\alpha$ generates corrections of order
$(\xi/\Xi)^{4\delta_\alpha}$ to observable quantities, where
$\delta_\alpha$ is the smallest scaling dimension of the
$f_\alpha$. For simplicity, we assume that $\xi \ll \Xi$; hence, since
we are seeking an upper bound on the computing time, we can ignore the
$\Gamma_\alpha$ corrections.

Another scenario to consider is when the physical qubits do not
interact with each other, $\xi \to \infty$. In this case, $a_{\beta
  jk} \to 0$ and the leading correction will come from
$\Gamma_\alpha$. This demands a slightly different organization of the
argument: It leads to a different definition of the effective coupling
constant but does not imply that there are no ``uncorrectable errors''
(see, e.g., Ref.\,\cite{PreskillNotes} for the stochastic error
model). Most of the following discussion can be readily adapted to
this case following arguments similar to those in Ref.\,\cite{NB06},
which we therefore omit here.

The steps outlined earlier result in the following quantum evolution
operator for the \emph{logical} qubits:
\begin{equation}
\bar{U}_I(T,0) \approx T_t\, e^{i \int_0^T dt \sum_{{\bf
      x},\alpha=\{x,z\}} \lambda_\alpha^\ast :f^\alpha ({\bf x},t):
  \bar{\sigma}_{\bf x}^\alpha}.
\label{eq:evol-coarse-grain}
\end{equation}
As a direct consequence of the coarse graining used in
Eq.\,(\ref{eq:evol-expanded}), note that the ultraviolet frequency
cutoff is $\Delta^{-1}$.

Equation (\ref{eq:evol-coarse-grain}) is a remarkable expression:
\emph{It shows that in the long wavelength limit the logical qubits
  obey the same dynamics as the physical qubits.} In other words, QEC
maps the ``spin-boson'' decoherence model onto itself. There are, of
course, several ways to reduce $\lambda_\alpha^\ast$: (i) engineer the
position of the physical qubits, (ii) change the distance of the code,
or (iii) concatenate the code. Nevertheless, as long as
$\lambda_\alpha^\ast$ and $\Gamma_\alpha$ are not strictly zero, there
will be degradation of the information in the logical qubits. Thus,
Eq.\,(\ref{eq:evol-coarse-grain}) implicitly defines the largest time
scale potentially available for computing.

\noindent
\emph{Upper-bound on the computational time}-- One way to quantify the
loss of quantum information to the environment is through the trace
distance \cite{NielsenChuang} between the reduced density matrix
$\rho_ R (T)$ and the ideal density matrix $\rho_0$: $D \big(
\rho_R(T),\rho_0 \big) = \frac{1}{2} \mbox{tr} \left| \rho_R (T) -
\rho_0 \right|$. The trace distance indicates how hard it is to
distinguish two density matrices by performing measurements; hence, it
is a natural way to quantify how well QEC protects information. Let us
suppose that there is a criterion $D \big( \rho_R(T), \rho_0 \big)
\leq D_{\rm crit}$ for a successful computation. Our goal, then, is to
evaluate the time $T$ available for computation.

Since we expect that $D \big( \rho_R(T), \rho_0 \big)$ is small, it is
natural to formulate the problem in powers of the effective couplings
$\lambda_\alpha^\ast$. For an upper bound on $T$, we can stop the
perturbative expansion in second order. Though it is difficult to
evaluate $D \big( \rho_R (T), \rho_0 \big)$ in general, we can make
some progress by considering two distinct cases. First, we look at an
isolated logical qubit, namely, $\Xi \to \infty$. Second, we use the
Hilbert-Schmidt norm to bound the trace distance and define an upper
bound on $T$ in general.

\noindent
\emph{Information lost by a single logical qubit}-- For a single
logical qubit, the trace distance can be expressed in terms of the
expectation values of the logical qubit $D \big( \rho_R (T), \rho_0
\big) = \sqrt{|\delta \sigma^+(T)|^2 + [\delta \sigma^z(T)]^2/4}$,
where $\delta \sigma^\alpha(T) = \langle\bar{\sigma}^\alpha (T)
\rangle - \langle \bar{\sigma}^\alpha \rangle$ and, for convenience,
we dropped the space label. Since the largest coupling constant is in
the $z$ direction, we employ a rotation to take it into account
nonperturbatively. First, we define the operator $:\!F^z
\big((n+1)\Delta\big)\!:-:\!F^z(n\Delta)\!:\, = \lambda_z^\ast
\Delta:\!f^z(n\Delta)\!:$ and then rotate the evolution operator at
each $n^{\rm th}$ QEC period using the unitary transformation $e^{-i\,
  :F^{z}(n\Delta):\, \bar{\sigma}^z}$. This rotation cancels the $z$
component of $H_I$ at the expense of dressing the transverse
coupling. The rotated interacting Hamiltonian at a time $t=n\Delta$
can be written as $ H_I^{\rm rot}(t) = \lambda_x^\ast
\sum_{\alpha=\{\pm\}} :f^\alpha(t): \exp\left[-2i\alpha
  :F^{z}(t):\right] \bar{\sigma}^\alpha.  $

We can now calculate the expectation values $\delta \sigma^\alpha(T)$
in perturbation theory in $\lambda_x^\ast$. This is a simple but
tedious calculation which we omit here. For our purposes, the main
feature appears already at zeroth order (dephasing only). In this
case, it is straightforward to show that $\left\langle
\bar{\sigma}^z(T) \right\rangle = \left\langle \bar{\sigma}^z
\right\rangle$ and $\left\langle \bar{\sigma}^+(T) \right\rangle =
e^{-4\gamma_z(T)} \left\langle \bar{\sigma}^+ \right\rangle$, where
$\gamma_z(T) = ( 2\pi/ L)^D (\lambda_z^\ast )^2 \sum_{{\bf k}\neq 0}
\frac{\left| u_{z,k} \right|^2} {\omega_{z,k}^2}
     [1-\cos(\omega_{z,k}T) ]$ is the well-known decoherence function
     \cite{breuer}. We thus obtain
\begin{equation}
      D \big( \rho_R (T),\rho_0 \big) = \left| \left\langle
      \bar{\sigma}^+ \right\rangle \right|\, \left[1 -
        e^{-4\gamma_z}(T) \right].
\label{eq:exact-distance}
\end{equation}
By defining $\zeta_z = 2(z_z - s_z) - D$, we can distinguish the
following decoherence regimes in the long-time limit:
\begin{equation}
\gamma_z(M\Delta) \propto \begin{cases} \left( \lambda_z^\ast/\omega_0
  \right)^2\, (\omega_0 \Delta)^{-\zeta_z/z_z}, & \zeta_z < 0,
  \\ \left( \lambda_z^\ast/\omega_0 \right)^2\, \ln M, & \zeta_z = 0,
  \\ \left( \lambda_z^\ast/\omega_0 \right)^2 (\omega_0
  \Delta)^{\zeta_z/z_z}\, M^{\zeta_z/z_z}, & 0 < \zeta_z < 2z_z,
  \\ (\lambda_z^\ast\Delta)^2\, \left( k_0L/2\pi
  \right)^{\zeta_z-2z_z} M^2, & \zeta_z > 2z_z, \end{cases}
\end{equation}
where $M\equiv T/\Delta$ is the number of QEC steps that are
performed. These regimes are straightforward generalizations of the
super-Ohmic ($\zeta_z<0$), Ohmic ($\zeta_z=0$), and sub-Ohmic
($\zeta_z>0$) regimes. Notice that for $\zeta_z < 0$, the trace
distance will converge to a finite value $D_{\rm sat}$. Equation
(\ref{eq:exact-distance}) is an exact result but we expect $D_{\rm
  crit}$ to be small. Hence, we can expand the exponential and find
the maximum time for computation with isolated logical
qubits. Assuming $D_{\rm crit} > D_{\rm sat}$, we find
\begin{equation}
M_{\rm max} \propto \begin{cases} \infty, & \zeta_z < 0, \\ \exp
  \left[ c_{D,z}\, D_{\rm crit}\, (\omega_0 / \lambda_z^\ast)^2
    \right], & \zeta_z = 0,
  \\ D_{\rm crit}^{z_z/\zeta_z}\,
  (\omega_0/\lambda_z^\ast)^{2z_z/\zeta_z}/(\omega_0\Delta), & 0 <
  \zeta_z < 2z_z, \\
  \left( 2 \pi / k_0 L \right) ^{\left( \zeta_z -2 z_z \right)}
   \sqrt{D_{\rm crit}} / (\lambda_z^\ast \Delta), &
  \zeta_z > 2z_z,
  \end{cases}
\end{equation}
where $c_{D,z}$ is a dimensionless prefactor of order unit.

\noindent
\emph{Upper bound for multiple logical qubits}-- To find an upper
bound on the trace distance when logical qubits are not isolated, we
use the sub-additivity property of the square root function and an
inequality proved in Ref.\,\cite{popescu},
\begin{equation}
D_{HS} \big(\rho_R (T),\rho_0 \big)
\leq D \big( \rho_R (T),\rho_0 \big) \leq 2^{\frac{N}{2}}
D_{HS} \big( \rho_R (T),\rho_0 \big),
\label{eq:DHSdef}
\end{equation}
where $D_{HS} \big( \rho_R (T),\rho_0 \big) = \frac{1}{2} [\mbox{tr} |
  \rho_R (T) - \rho_0 |^{2}]^{1/2}$ is the Hilbert-Schmidt norm and
$N$ is the number of logical qubits. Following a similar procedure to
that used for the trace distance, we can expand $D_{HS} \big( \rho_R
(T),\rho_0 \big)$ to second order in $\lambda_\alpha^\ast$,
\begin{equation}
\label{eq:DHS}
D_{HS} \big( \rho_R (T),\rho_0 \big) \propto \sqrt{\sum_\alpha
  (\lambda_\alpha^\ast)^2 \Big| \sum_{{\bf x},{\bf y}} W_{{\bf x},{\bf
      y}}^\alpha (T) \Big|^{2}},
\end{equation}
\begin{equation} 
W_{{\bf x},{\bf y}}^\alpha (T) = \left( \frac{2\pi}{L} \right)^D
\sum_{{\bf k}\neq 0} \frac{\left| u_{\alpha,k} \right|^2}
    {\omega_{\alpha,k}^2} e^{-i {\bf k} \cdot ({\bf x}-{\bf y})}
    \!\!\left( 1- e^{-i\omega_{\alpha,k}T} \right)\!.
\end{equation} 
There are two types of $W^\alpha_{{\bf x},{\bf y}}(T)$: (i) the
diagonal self-interaction terms, and (ii) the correlation terms in
which pairs of logical qubits interact. Both types lead to the same
functional dependence in the sum:
\begin{equation}
\label{eq:Marray}
  \Big| \sum_{{\bf x},{\bf y}} W_{{\bf x},{\bf y}}^\alpha (T) \Big|
  \propto \begin{cases} N \omega_0^{-1}\, (\omega_0
    \Delta)^{-\zeta_\alpha/z_\alpha}, & \zeta_\alpha < 0, \\ N
    \omega_0^{-1}\, \ln M, & \zeta_\alpha = 0, \\ N \omega_0^{-1}
    (\omega_0 \Delta\, M)^{\zeta_\alpha/z_\alpha}, & 0<\zeta_\alpha <
    z_\alpha, \\ N \Delta\, \left( k_0L/2\pi
    \right)^{\zeta_\alpha-z_\alpha} M, & \zeta_\alpha >
    z_\alpha, \end{cases}
\end{equation} 
where the proportionality constant is of order 1. However, the two
types of terms lead to different onset criteria. For the
self-interacting part, the different regimes are delineated using
$\zeta_\alpha = 2(z_\alpha - s_\alpha) - D$, while for the correlation
part, the spatial sum leads to $\zeta_\alpha = 2 (z_\alpha - s_\alpha)
+ D_x - D$ with $D_x$ being the dimension of the qubit array. Note
that some physical arrangements of qubits are more favorable than
others; for instance, a linear architecture is more favorable than a
square or cubic one.

First, note that in order to apply QEC we assumed ${\lambda^{*}}^2 N
\ll 1$ [Eqs.\ (\ref{eq:DHS}) and (\ref{eq:Marray})]. Second, for a
given critical distance $D_{\rm crit}$ and using
Eq.\,(\ref{eq:Marray}), we arrive at an upper bound on the time
available to compute due to each component of the environment:
\begin{equation}
M_{\rm max} = \begin{cases} \infty, & \zeta_\alpha < 0, \\ \exp \left[
    \frac{b_{D,\alpha}\, D_{\rm crit}}{ N (\lambda_z^\ast/\omega_0)}
    \right], & \zeta_\alpha=0, \\ (\omega_0\Delta)^{-1} \left[
    \frac{D_{\rm crit}}{ N (\lambda_z^\ast/\omega_0)}
    \right]^{z_\alpha/\zeta_\alpha}, & 0 < \zeta_\alpha < z_\alpha,
  \\ \left( 2\pi/k_0L \right)^{\zeta_\alpha - z_\alpha} \frac{D_{\rm
      crit}}{ N (\lambda_z^\ast\Delta)}, & \zeta_\alpha >
  z_\alpha, \end{cases}
\end{equation} 
where $b_{D,\alpha}$ is a dimensionless constant. If $\lambda_x \sim
\lambda_z \sim \lambda$, this result is simply related to the code
distance or the level of concatenation: $\lambda_\alpha^\ast \sim
\lambda^{d_c}$.

\noindent
\emph{Conclusions}-- For how long is it possible to quantum compute?
(1) The trace distance calculations give us a rule of thumb: For a
finite computation time, the residual decoherence of a logical qubit
after the first QEC step times the number of logical qubits must be a
small number, $(\lambda_\alpha^\ast)^2 N \ll 1$. (The case of no
spatial correlation among the physical qubits at short times,
$\lambda_\alpha^\ast = 0$, was discussed in Ref.\,\cite{NB06}.) This
condition must be a factor in the choice of the distance of the code
or concatenation level. (2) While the argument presented here does not
directly address the threshold theorem, the upper bound on the
available computational time shows that there are certain limits to
the power of QEC. The three regimes that we find nicely fit the
qualitative interpretation of resilience as a dynamical quantum phase
transition \cite{NB06}. (2.1) For $\zeta_{x,z} < 0$ (above the ``upper
critical dimension''), the usual enunciation of the threshold theorem
\cite{B-T} can be used, and therefore it is possible to compute
indefinitely. (2.2) For $\zeta_{x,y}>z_{x,y}$ (below the ``lower
critical dimension''), correlations are so strong that the available
computational time is formally zero (since it depends on the size of
the bath, $L$). It is, however, conceivable that its strong infrared
divergence may be handled by combining dynamical decoupling and QEC
methods \cite{DD+FT}. (2.3) Finally, between these two regions, there
is a maximum time available to compute. This constraint must also be a
factor in the choice of the distance of the code or concatenation
level. Even though the regimes fit into the general discussion of
Ref.\,\cite{NB06}, the definition of the upper critical dimension
given here is not the same. The reason is that, we have now shown that
it is possible to \emph{explicitly} treat a dense set of qubits.

\vspace{-.1cm}

We thank D. Lidar for useful discussions and correspondence. This work
was partially supported by the Office of Naval Research and
CNPq-Brazil.



\end{document}